\documentclass[aps,prb,twocolumn,groupedaddress,showpacs]{revtex4-1}

\usepackage{amsmath}
\usepackage{graphicx}
\usepackage{dcolumn}

\begin{document}
\title{Magnetic blockade mechanism for quantum nucleation of superconducting vortex-antivortex pairs in zero external magnetic field}

\author{J.~H.~Miller,~Jr.}
\email[]{jhmiller@uh.edu}
\affiliation{Department of Physics, University of Houston, Houston, Texas 77204-5005 USA}
\affiliation{Texas Center for Superconductivity, University of Houston, Houston, Texas 77204-5002 USA}
\author{A.~I.~Wijesinghe}
\affiliation{Department of Physics, University of Houston, Houston, Texas 77204-5005 USA} 
\affiliation{Texas Center for Superconductivity, University of Houston, Houston, Texas 77204-5002 USA}

\date{\today}

\begin{abstract}
We propose a magnetic dual of the Coulomb blockade effect for quantum nucleation of flux vortex pairs in high-$T_c$ superconducting (HTS) films and grain boundaries in zero applied field. The magnetic blockade instability occurs at $\theta=\pi$, where $\theta$ is the ``vacuum'' or theta angle. The $\theta$ term has recently been discussed in the context of  several other systems, including charge and spin density waves, topological insulators, the quantum Hall effect, and spontaneous CP violation. Our model predicts a sharp pair creation threshold current at $\theta=\pi$, analogous to the Coulomb blockade voltage of a tunnel junction, and explains the observed thickness dependence of critical currents in HTS coated conductors. We use the Schr\"{o}dinger equation to compute the evolving macrostate amplitudes, coupled by a generalized tunneling matrix element. The simulations yield excellent quantitative agreement with measured voltage-current characteristics of bi-crystal and other HTS grain boundary junctions. The model also predicts non-sinusoidal behavior in the voltage oscillations resulting from time-correlated vortex tunneling.
\end{abstract}


\maketitle
\section{INTRODUCTION}
The last several decades have seen a crumbling of the barrier between the quantum and macroscopic realms. High-$T_c$ superconductivity provides an exquisite example of long range quantum coherence at temperatures above~77~K. Cooperative quantum tunneling has also emerged as an important class of phenomena, whose manifestations include coherent Josephson tunneling~\cite{Josephson} and tunneling of vortices,~\cite{3,4,5} which can occur more readily in layered high-$T_c$ superconductors due to their pancake-like structures.~\cite{E.M.Chudnovsky}

Vortices are known to nucleate in superfluid helium above a critical velocity,~\cite{Ihas1992} underscoring the point that magnetic flux is not crucial to vortex nucleation. In fact, as will be discussed, the magnetostatic energy contained within a superconducting flux vortex actually impedes vortex nucleation for applied currents below a critical value. Both Abrikosov and Josephson vortices, the latter being especially important in high-$T_c$ superconducting~(HTS) grain boundaries, play important roles in limiting the critical currents of HTS films and tapes. A Josephson vortex is a 2$\pi$ flux soliton in a wide Josephson junction (wide compared to the Josephson penetration length~$\lambda_J$), whose energy includes the extra Josephson coupling energy in the region where the phase difference~$\phi$ across the junction advances by~2$\pi$ from one minimum to the next. 

Though spatially extended, Josephson vortices are extremely light. Using the expression provided by Grosfeld and Stern,~\cite{E.Grosfeld} the effective mass of a Josephson vortex spanning a 1-$\mu$m thick film is estimated to be $\sim10^{-2}m_e$. The mass of a single pancake Josephson vortex (Josephson pancake~\cite{R.G.Mints}) within the $k^{th}$ superconducting cuprate or pnictide layer, whose phase difference~$\phi_k$ advances by 2$\pi$,  is several orders of magnitude smaller still.  An applied current in zero external magnetic field can induce a vortex-antivortex pair to nucleate in close proximity by quantum tunneling~\cite{A.Widom} and subsequantly expand outward, as illustrated in Fig.~\ref{fig1}(a). 

Conceptually, this process is analogous to electron-hole pair creation via Landau-Zener tunneling through a tilted bandgap in a semiconductor or Schwinger pair production of electron-positron pairs out of the vacuum.~\cite{T.D.Cohen} In this case, the applied current~$I$ acts as the driving force. Neglecting the magnetostatic energy and using the expression derived by Cohen and McGady~\cite{T.D.Cohen} for the pair production rate, one finds the voltage across a wide grain boundary~JJ to be proportional to $I$~exp[-$I_0/I$]. Although highly nonlinear, this lacks a sharply defined threshold current below which the voltage is identically zero. However, recent experiments probing HTS bi-crystal grain boundaries with femtovolt precision~\cite{Kalisky2009} suggest the existence of a sharp critical current below which no Josephson vortices nucleate, either in the middle or at the edges of the grain boundary.

\section{MAGNETIC BLOCKADE THRESHOLD CURRENT FOR VORTEX NUCLEATION}

The Coulomb blockade effect is well known for single electron tunneling,~\cite{D.V.Averin} and has also been proposed for charge soliton pair creation in the massive Schwinger model,~\cite{S.Coleman.1976} essentially a quantum sine-Gordon model that includes electrostatic effects, and in density waves.~\cite{I.V.Krive,J.H.Miller,J.H.Miller.2011} Here we propose a \textit{magnetic} blockade threshold current for Abrikosov or Josephson vortex pair creation~\cite{A.Widom} in HTS films and grain boundaries.~\cite{H.Hilgenkamp} This threshold, for Josephson vortex nucleation in a grain boundary, can be much smaller than the Ambegaokar--Baratoff critical current even when \textit{d}--wave pairing symmetry is taken in to account.~\cite{J.H.Xu} A magnetic ``Weber" blockade effect has also recently been proposed to interpret magneto-conductance oscillations of thin, narrow superconducting strips.~\cite{D.Pekker} 

\begin{figure}[ht] 
\includegraphics[scale=.5]{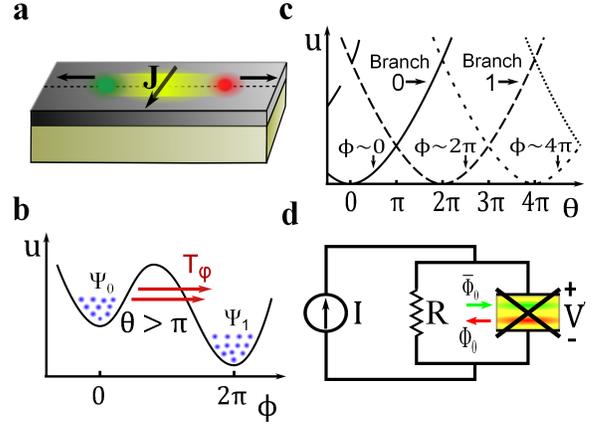}
\caption{\textbf{(a)}.~Nucleation of a Josephson vortex-antivortex pair (green and red) in an HTS grain boundary. The yellow shading indicates the magnetic flux linking the vortices. \textbf{(b)}.~$u(\phi)$ when  $\theta=2\pi E/E^*> \pi$ as the phases  $\phi_k(x)$ tunnel coherently into the lower well. \textbf{(c)}.~Potential energy~ vs.~$\theta$   for  $\phi \sim2\pi n$. \textbf{(d)}.~Modified resistively shunted junction model, in which time-correlated tunneling of Josephson vortices and antivortices is represented by analogy to time-correlated SET.~\cite{D.V.Averin}\label{fig1}}
\end{figure}

Figure~\ref{fig1}(a) illustrates nucleation, in zero applied magnetic field, of a pair of Josephson vortices in a wide HTS grain boundary.  The vortex pair generates a magnetic field (yellow shading): $B^*=\beta\Phi_0/\lambda^2$ that links the vortex and the anti-vortex. Here $\lambda=\sqrt{\lambda_J\lambda_L}$  or $\lambda=\lambda_L$ for a Josephson or Abrikosov vortex pair, respectively, $\Phi_0=h/2e$ is the flux quantum, $\lambda_L$ is the London penetration length, and  $\beta\sim 1/\pi$ is a geometrical factor. A current density~$J$ in a film of thickness~$d$ creates a magnetic field, $B_J\approx\mu_0Jd/2$, which can partially cancel the internal field linking the vortex and antivortex and thus overcome the magnetic barrier (Coulomb barrier analog), provided its value is high enough. 

At low applied currents, the difference in magnetostatic energy densities with and without the vortex pair, $(B_J\pm B^*)^2/2\mu_0-B_J^2/2\mu_0$, is positive when $\left|B_J\right|< B^*/2$. However, when $\theta=2\pi B_J/B^* > \pi$, the formerly lowest energy state becomes a metastable state or ``false vacuum,'' as illustrated in Fig.~\ref{fig1}(b). The phases $\phi_k$ tunnel coherently into the lower energy well, creating a bubble of ``true vacuum'' bounded by flux solitons (Josephson vortices) when~$\theta$  exceeds~$\pi$. Setting $\theta~=~\pi$ thus yields the pair creation threshold current per unit width:
\begin{equation}
j_{pc}^0=B^*/\mu_0 = \beta\Phi_0/(\mu_0\lambda^2).\label{eq:eq0}
\end{equation}

For the case of Josephson vortex nucleation, where  $\lambda^2 = \lambda_L \lambda_J$, this yields a scaling with classical critical current density $J_0$ of the form, $j_{pc}^0 \propto  J_0^{1/2}$, since $ \lambda_J \propto  J_0^{-1/2}$ for both planar and sandwich-type junctions.~\cite{AppPhyLett69.3914}
For currents below the vortex nucleation threshold, $\theta$  can be written in terms of the supercurrent per unit width  $j_s$ as  $\theta=\pi\left(j_s/j_{pc}^0\right)$. If the vortex and antivortex are separated by a distance~$\ell$, then the pair creation critical current between vortices can be written as $j_{pc}^0\ell\sim\Phi_0/2L$, where $L\sim\mu_0\lambda^2/2\beta\ell$ is the inductance of the flux toroid coupling the vortices. This is essentially the dual of the Coulomb blockade voltage $V_c = e/2C$ for a small capacitance tunnel junction.~\cite{D.V.Averin} Nucleation of a vortex near one edge and antivortex near the other can be treated similarly using the image vortex concept.~\cite{A.A.Abrikosov}

HTS coated conductors generally have small-angle grain boundaries~\cite{H.Hilgenkamp} and their critical currents vs. thickness~$d$ often plateau, so that~$J_c$ decreases with $d$.~\cite{S.R.Foltyn} This is readily explained by the vortex pair creation current density, $J_{pc}^0=\beta\Phi_0/(\mu_0\lambda^2d) $ when $d > \lambda$, resulting from the magnetic blockade effect.  For a strip of width $w$, the pair creation current becomes: $I_{pc}^0=\beta\Phi_0w/(\mu_0\lambda^2)$, which is independent of $d$ until it becomes large enough for vortex rings to nucleate. When $d <<\lambda$, the size of the vortices increases since the effective 2-D penetration length~$\Lambda$  scales inversely with~$d$:~\cite{J.Pearl} $\Lambda=\lambda^2/d$. Since $\Lambda=\lambda$ when $d>>\lambda$, we use the approximation: $\Lambda\approx\lambda+\lambda^2/d$, which yield the vortex pair creation critical current:
\begin{equation}
I_{c}=I_{pc}(d)=\frac{I^0_{pc}}{\left[1+\lambda/d\right]^2}.\label{eq:eq1}
\end{equation}

\begin{figure}[hb] 
\includegraphics[scale=.5]{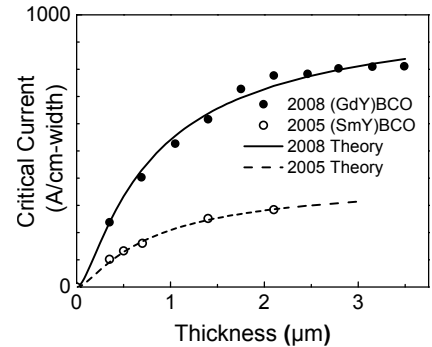}
\caption{Theory vs. experiment~\cite{V.Selvamanickam} for critical currents of HTS coated conductors vs. thickness. (See text: we use  $\lambda$= 380~nm, $I_{pc}^0$= 1030~A and 398~A to fit 2008 and 2005 data, respectively.)\label{fig2}}
\end{figure}

Figure~\ref{fig2} shows a favorable comparison of Eq.~(\ref{eq:eq1}) with measured critical currents~vs.~thickness of HTS coated conductors. The introduction of multiple insulating CeO$_2$ spacer layers into HTS coated conductors has been found to increase critical current.~\cite{S.R.Foltyn} Our model provides a straightforward interpretation~--~sufficiently thick spacer layers decouple vortex pair nucleation events on adjacent HTS layers, enabling~$I_c$ to increase with the number of layers.

\section{TIME CORRELATED VORTEX TUNNELING: THE SCH\"{o}DINGER EQUATION AS AN EMERGENT CLASSICAL DESCRIPTION}

The ``vacuum'' or theta angle~$\theta$, discussed extensively in the quantum field theory literature (e.g. see Ref.~\cite{S.Coleman.1976} and citing papers) is related in our model to total displacement flux~$\Phi$  by: $\theta=2\pi\left(\Phi/\Phi_0\right)$, where $\Phi=\lambda^2B_J/\beta+n\Phi_0$. The potential energy of the $k^{th}$ layer for a cuprate or pnictide superconductor can then be written, similar to the massive Schwinger model,~\cite{S.Coleman.1976} as:
\begin{equation}
u[\phi_k]=u_{_J}\left[1-cos\phi_k(x)\right]+u_{_M}\left(\theta-\phi_k(x)\right)^2~,\label{eq:eq2}
\end{equation}
where the first term is the Josephson coupling energy and the quadratic term is the magnetostatic contribution. 

Figure~\ref{fig1}(c) shows plots of $u$~vs.~$\theta$   when the energy is minimized for $\phi\sim2\pi n$ (dropping the subscript~$k$) when $u_M<<u_0$. Each $u~vs.~\phi$ branch has a parabolic form, $\propto~(\theta~-~2\pi n)^2 \propto~(\Phi~-~n\Phi_0)^2/2L$, analogous to the charging energy parabolas, $\propto~(Q~-~ne)^2/2C$, in time-correlated SET~\cite{D.V.Averin} and time-correlated soliton tunneling.~\cite{J.H.Miller,J.H.Miller.2011} The phases~$\phi_k$ tunnel coherently into the next well as each parabola, or branch, in Fig.~\ref{fig1}(c) crosses the next at $\theta=\pi$ and the other instability points $\theta = 2\pi \left(n+1/2\right)$.  Regardless of the detailed shape of the Josephson coupling energy (which may or may not be sinusoidal and may even include disorder), the behavior that emerges is clearly nonsinusoidal in nature, which affects the Shapiro steps in the $I-V$ curve in the presence of a microwave signal.

The approach proposed here is based on the hypothesis that the amplitudes $\psi_n$ and  $\psi_{n+1}$ for the system to be on branches $n$ and $n+1$, respectively, are coupled via coherent tunneling of Josephson pancakes, represented as microscopic quantum solitons.~\cite{A.Maiti,Z.Hermon} This picture is motivated by Feynman's intuitive derivation~\cite{R.P.Feynman} of the dc and ac Josephson effects. Advancing  $\phi_k(x)$ by $2\pi$ in a finite region is equivalent to creating a pair of microscopic solitons~--~essentially pancake Josephson vortices~\cite{R.G.Mints} with extremely small effective masses.~\cite{E.Grosfeld} Similarly, Abrikosov pancake vortices~\cite{A.Grigorenko} are pointed out to have small effective masses.~\cite{E.M.Chudnovsky} In this picture, many coupled pancake vortices behave as a quantum fluid,~\cite{prl77.3901} within which their coherent tunneling is viewed as a secondary Josephson effect. 
 
We use a modified version of the resistively shunted junction~(RSJ) model [Fig.~\ref{fig1}(d)], in which the analogy to time-correlated single electron tunneling is used to simulate time-correlated vortex tunneling and obtain voltage-current characteristics of an HTS grain boundary junction. The voltage across the junction is: $V=d\Phi/dt=(\Phi_0/2\pi) d\theta/dt$, yielding a normal current through the shunt resistance: $I_n=V/R=(\Phi_0/2\pi R) d\theta/dt$. For a grain boundary junction with total current~$I$, we define $\omega=(2\pi R/\Phi_0)I$. 

In a wide junction, advancing the phase by  $\left\langle \phi\right\rangle$ (the average phase for all layers) near the middle of the junction creates a Josephson kink-antikink pair whose circulating currents either reduce or reinforce the existing supercurrent between kinks. This yields a net supercurrent given by: $I_s=(I^*/2\pi)\left[\theta-\left\langle \phi\right\rangle\right]$. Using $I_n=I-I_s$ and introducing the time constant $\tau=L/R$, yields the following equation for the time evolution of~$\theta$: 
\begin{equation}
\frac{d\theta}{dt}=\omega-\frac{1}{\tau}\left[\theta-\left\langle \phi \right\rangle\right].\label{eq:eq3}
\end{equation}
The expectation value $\left\langle \phi\right\rangle$  is computed by solving the time-dependent Schr\"{o}dinger equation:
\begin{equation}
i\hbar\frac{\partial \psi_{0,1}}{\partial t}=U_0\psi_{0,1}+T_{\varphi}\psi_{1,0}~,\label{eq:eq4}
\end{equation}
viewed as an emergent classical equation following Feynman,~\cite{R.P.Feynman} to compute the original and emerging macrostate amplitudes $\psi_0(t)$  and $\psi_1(t)$. These are coupled via a tunneling matrix element~$T_{\varphi}$ with a Zener-like force dependence, which transfer the phase $\phi_k$ from one well to the next [Fig.~\ref{fig1}(b)].

Our model represents the amplitudes $\psi_{ 0,1}$ for the system to be on branches 0 and 1 [Fig.~\ref{fig1}(c), more generally $n$ and~$n+1$] by:
\begin{equation}
\psi_{0,1}=\sqrt{\rho_{0,1}}\exp{i\delta_{0,1}}~.\label{eq:eq5}
\end{equation}
For a layered superconductor,  $\rho_{0,1}=N_{0,1}/N$ is the fraction of superconducting layers on the respective branch. In this picture $\psi_{0,1}$ are $not$ this superconducting order parameters on opposite sides of the junction but, rather, represent the amplitudes for the phases $\phi_k$ to be in either of the two wells shown in Fig.~\ref{fig1}(b). The phases $\delta_{0,1}$ are the Berry phases for the vortices, which would, for example, lead to an Aharonov-Casher effect for vortices traveling along two branches of a ring-shaped Josephson junction surrounding an island charge.~\cite{Elion1993}

Advancing  $\phi_k(x)$ by 2$\pi$  within a given region, taking  $\phi_k$ from one branch to the next, is equivalent to creating a pair of microscopic 2$\pi$-solitons (Josephson vortices). The driving force is the energy difference per unit length between potential minima at  $\phi\sim2\pi n$ and   $\phi\sim2\pi(n+1)$. When $u_M/u_0<<1$, this force is: $F=4\pi u_M{\theta_n}'$, where ${\theta_n}'\equiv\theta-2\pi(n+\frac{1}{2})$. 

Following Bardeen's procedure,~\cite{J.Bardeen,C.B.Duke} the matrix element for Zener tunneling through the soliton energy gap is estimated as: 
\begin{equation}
T_{\varphi}(F)=-4F\lambda\exp[-F_0/F]~,\label{eq:eq6}
\end{equation}
where $\lambda^{-1}\sim\Delta_\varphi/\hbar v_0+\lambda_m^{-1}$,   $\lambda_m$ is a mean free path length,  $\Delta_\varphi$  is the microscopic soliton energy for a pancake Josephson vortex, $v_0$ is the phase (Swihart) velocity,~\cite{J.Swihart} and $F_0\sim\Delta_\varphi^2/\hbar v_0$. This expression for $T_\varphi$ is similar to the rate of Zener tunneling or Schwinger pair production in 1-D.~\cite{T.D.Cohen} Since any negative energy difference within the ``bubble'' is balanced by the positive flux soliton pair energy, the matrix element couples states of equal energy, $U_0=U_1=U$. Thus, defining  $\psi_{0,1}=\chi_{0,1}(t)\exp[-iUt/\hbar]$, the Schr\"{o}dinger equation [Eq.~(\ref{eq:eq4})] reduces to:  
\begin{equation}
i\hbar\partial\chi_{0,1}/\partial t=T_{\varphi}\chi_{1,0}.\label{eq:eq7}
\end{equation}

Below the pair creation threshold, we can write: $\theta=2\pi I/I^*$  where $I^*=2\beta \Phi_0 \ell/(\mu_0 \lambda^2)=(\Phi_0/L)(\ell/w)$ within a grain boundary region of width $\ell$, which may be slightly smaller than $w$ due to the finite width of the vortices. Due to multiple vortex pair nucleation events in a wide junction, $\ell/w$  may not decrease monotonically with $w/\lambda_J$. The variables are put into dimensionless form: $t'\equiv t/\tau$, $f\equiv\omega\tau/2\pi$, proportional to total current $I$, $q\equiv\theta/2\pi$, where  $q_0\equiv \theta_0/2\pi=F_0/(4\pi u_M )$, ${q_n}'\equiv{\theta_n}'/2\pi=q-n-1/2$, and $i_s\equiv I_s/I^*= q-p-n$  is the normalized supercurrent, and the normalized voltage is: $v\equiv\frac{dq}{dt'}=f-i_s$. Finally, setting $\chi_0(t)=c_0(t)$ and $\chi_1(t)=ic_1(t)$ in Eq.~(\ref{eq:eq7}), taking $c_0$ and $c_1$ to be real, yields the following coupled differential equations: 
\begin{equation}
\begin{aligned}
\frac{dc_1}{dt'}&= ~~[\gamma {q_n}'\exp(-q_0/{q_n}')] c_0~,\\
\frac{dc_0}{dt'}&= -[{\gamma q_n}'\exp(-q_0/{q_n}')] c_1~,
\end{aligned}
 \label{eq:eq8}
\end{equation}
for $q_n'~>~0$ and where  $\gamma=32\pi^2u_M\lambda$. These are integrated numerically with initial values $c_0 =1$ and $c_1=0$. The phase expectation value is: $\left\langle \phi\right\rangle=2\pi[n+p]$, where $p=|c_1|^2$, and the transition is considered complete when $p$ exceeds a cutoff close to one. 

For each value of $f=I/I^* $, a time average over several complete cycles is performed to compute~$v=(\ell/w)(V/I^* R)$.  A similar approach has recently been used by the authors to compute current-voltage characteristics due to soliton pair production in density waves.~\cite{J.H.Miller.2011}  As in the density wave case, each transition from one branch to the next takes place over a relatively long time interval, suggesting that many pancake vortices ``flow'' through the barrier like a quantum fluid rather than tunneling abruptly as a single massive flux line. 

\begin{figure}[ht] 
\includegraphics[scale=.65]{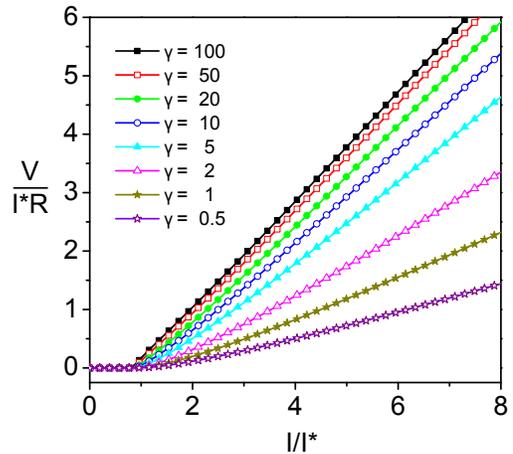}
\caption{(Color online.) Normalized theoretical voltage-current plots for a fixed value of $q_0 = 3$ and several values of $\gamma$.\label{fig22}}
\end{figure}

Figure~\ref{fig22} shows resulting theoretical normalized voltage vs. current plots for several values of  $\gamma$ for a fixed value of $q_0 = 3$ and $\ell/w=1$.  The \textit{V-I} plots exhibit piecewise linear behavior, which fits neither the classical RSJ model nor a straightforward thermally activated flux flow model. Similar piecewise quasi-linear behavior is seen quite frequently in high-$T_c$ superconducting YBCO grain boundary junctions. The main effect of increasing $\gamma$  is to increase the slope of the \textit{V-I} curve, saturating at a slope of one for the normalized plot.

\begin{figure}[ht] 
\includegraphics[scale=.65]{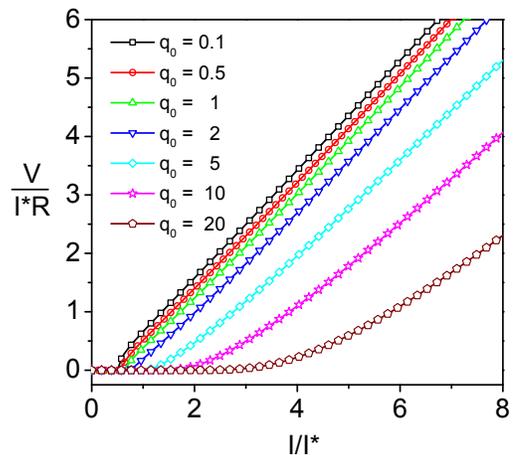}
\caption{(Color online.) Normalized simulated voltage-current plots for a fixed value of  $\gamma = 20$ and several values of  $q_0$.\label{fig23}}
\end{figure}

Figure~\ref{fig23} shows similar plots for a fixed value of  $\gamma = 20$ and several values of $q_0$.  Here, the main effect of increasing $q_0$, which corresponds to increasing the energy required to nucleate each vortex-antivortex pair, is to increase the degree of rounding in the \textit{V-I} plot. Moreover, note that the simulated ``measured'' critical current $I_c$ increases to well above the nominal magnetic blockade critical current~$I_{pc} \cong I^*/2$ as $q_0$ becomes significantly greater than one. 

\begin{figure}[ht] 
\includegraphics[scale=.6]{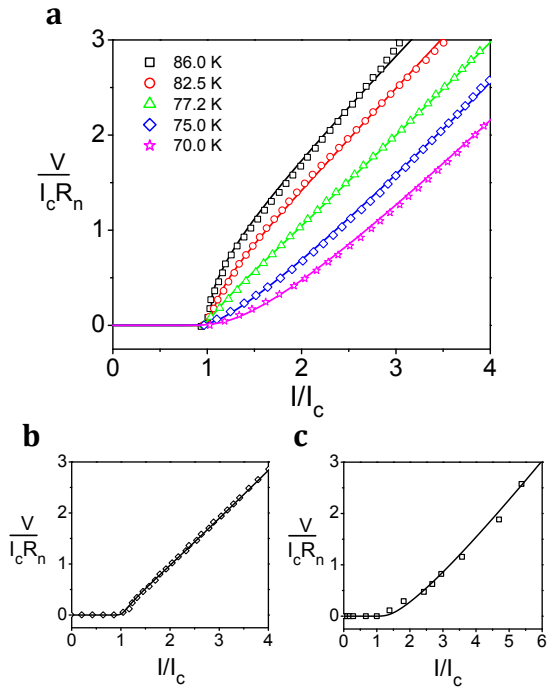}
\caption{(Color online.) \textbf{(a)}.~Experiment vs. theory for a YBCO bicrystal junction.~\cite{R.D.Redwing} Top curves: \textit{V-I} curve at 86~K (dotted line) vs. classical RSJ model (solid line).  Remaining curves: quantum nucleation model (solid lines) vs. measured \textit{V-I} curves (dotted lines) at (top to bottom): 82.5~K, 77.2~K, 75.0~K, and 70.0~K. \textbf{(b)}.~Comparison between quantum nucleation model (solid line) vs. \textit{V-I} characteristic (diamonds) of an iron pnictide superconductor bicrystal at 4.2~K.~\cite{X.Zhang} \textbf{(c)}.~Model simulation vs. experiment for \textit{V-I} curve of grain boundary junction in a thallium-based cuprate superconducting film at 77~K.~\cite{C.Dark}\label{fig3}}
\end{figure}

Figure~\ref{fig3}(a) shows comparisons between theory and the measured~\textit{V-I} characteristics of a YBCO~\cite{R.D.Redwing} grain boundary junction at several temperatures. The simulated 86~K plot in Fig.~\ref{fig3}(a)~(top) was obtained using the classical RSJ model in the overdamped limit: $V/(I_c R_n) = \sqrt{(I/I_c )^2-1}$, without invoking thermal activation. The fact that the 86~K data fits the classical RSJ model almost perfectly suggests that the effective Josephson penetration length is comparable to or longer than the junction width (short junction limit) due to the small Josephson coupling energy at this temperature. 

The theoretical plots for the remaining temperatures in Fig.~\ref{fig3}(a), as well as for Figs.~\ref{fig3}(b) and~\ref{fig3}(c), were obtained from the Schr\"{o}dinger equation as discussed above, using the parameters shown in Table~\ref{table1}. The increase in $q_0$ going from 82.5~K down to 70.0~K is consistent with increasing Josephson coupling energy, leading to a larger $F_0$ and smaller $\lambda_J$, as the temperature decreases.  The increased rounding of the \textit{V-I} curves with decreasing $T$ is also consistent with the system going from the short- to the long-junction limit as  $\lambda_J$ decreases. Moreover, the fact that the \textit{V-I} curves are more rounded for low than for high temperatures provides powerful evidence that the rounding of the \textit{V-I} curves is \textit{not} due to thermal activation. The extremely light masses of Josephson vortices coupled with quantum fluidic properties, discussed above, enable quantum effects to dominate over the entire temperature range. Further evidence that even Abrikosov vortices exhibit quantum, rather than thermally activated, behavior over a wide temperature range is provided by magnetic relaxation rates of trapped flux in YBCO, which go \textit{down} with increasing temperature~\cite{Y.Y.Xue} from 4~K to  about~86~K.

Figure~\ref{fig3}(b) shows excellent agreement between our quantum model and the \textit{V-I} curve of an iron pnictide superconducting bicrystal,~\cite{X.Zhang} consisting of coupled SrFe$_{1.74}$Co$_{0.26}$As$_2$ and Ba$_{0.23}$K$_{0.77}$Fe$_2$As$_2$ crystals, each $\sim$300~$\mu$m wide, and thus in the long junction limit. Similar piecewise linear behavior to that seen in the figure also occurs frequently in cuprate grain boundary junctions [e.g. Fig.~\ref{fig3}(a), 77.2~K data and Ref.~\cite{D.Dimos,F.Lombardi}] and is analogous to the piecewise linear \textit{I-V} curve of an ideal Coulomb blockade tunnel junction.  Figure~\ref{fig3}(c) shows a plot of our simulation as compared to the measured \textit{V-I} characteristic of a thallium-based cuprate,~\cite{C.Dark} exhibiting good agreement with the data within experimental error. Table~\ref{table1} shows the parameters used for the simulations. 

\begin{table}[h]
\caption{\label{table1}Parameters used to generate the simulated \textit{V-I} curves in~Fig.~\ref{fig3}.}
\begin{ruledtabular}
\begin{tabular}{ccccc}
\textrm{Figure Plot}&
\textrm{$I_c/I^*$}&
\textrm{$w/\ell$}&
\textrm{$\gamma$}&
\textrm{$q_0$}      \\
\colrule
Fig.~\ref{fig3}(a) - 82.5~K&0.55 &1.15 &50 &0.5 \\
Fig.~\ref{fig3}(a) - 77.2~K&0.72 &1.09 &45 &1.7 \\
Fig.~\ref{fig3}(a) - 75.0~K&1.25 &1.16 &40 &6.5 \\
Fig.~\ref{fig3}(a) - 70.0~K&1.80 &1.16 &30 &12.0 \\
Fig.~\ref{fig3}(b)         &0.71 &1.04 &60 &1.9 \\
Fig.~\ref{fig3}(c)         &1.00 &1.00 &12 &6.0 \\
\end{tabular}
\end{ruledtabular}
\end{table}

\begin{figure}[h] 
\includegraphics[scale=.50]{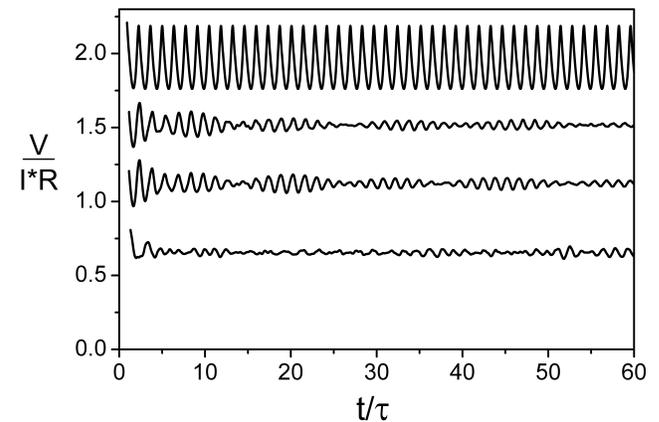}
\caption{~Simulated normalized voltage oscillations vs. normalized time for (top-to-bottom, offset for clarity): a uniform junction with $\gamma=40$ and $q_0=6.5$; and nonuniform junctions, represented as 100 junctions in parallel using the same average $\gamma$ and $q_0$, but (see text) with $\delta=0.1,~k=0.05$, $\delta=0.1,~k=0.1$, and $\delta=0.5,~k=0.1$. \label{fig4}}
\end{figure}

The use of the time-dependent Schr\"{o}dinger equation and generalized tunneling matrix element $T_\varphi$, coupled with the analogy to time-correlated single electron tunneling suggested in Fig.~\ref{fig1}, provides a simple, yet powerful approach to modeling the dynamics of vortex tunneling. The top plot in Fig.~\ref{fig4} shows predicted voltage oscillations, exhibiting highly non-sinusoidal behavior, for a uniform junction with the parameters $\gamma=40$ and $q_0=6.5$, for an applied current $f=I/I^*=2.5$. 

In the remaining plots of Fig.~\ref{fig4} we model non-uniformities by representing the junction as 100 junctions in parallel, for which $\gamma$ and $q_0$ are psuedo-randomly varied with a uniform distribution within the ranges $\gamma(1\pm\delta)$ and $q_0(1\pm\delta)$, respectively. The phases between adjacent domains $i$ are coupled via an additional term that adds to the net force for each domain: 
$k\left[\left\langle \phi-{i+1}\right\rangle-2\left\langle \phi_i\right\rangle\left\langle \phi_{i-1}\right\rangle\right]$, applying periodic boundary conditions. 
The remaining plots in Fig.~\ref{fig4} also assume $f=2.5$ and the same average values for $\gamma$ and $q_0$ as before, but with different values of $\delta$ and $k$. The middle two plots, for which $\delta=0.1$ and $k=0.05$ (second from top) and $k=0.1$, show greatly reduced voltage oscillation amplitudes as well an apparent amplitude modulation effect consistent with reports of subharmonic Shapiro steps.~\cite{Terpstra1995} The bottom plot ($\delta=0.5$ and $k=0.1$), representing the highest degree of disorder for this series, shows oscillation amplitudes that are even further reduced. This is consistent with the idea that Shapiro steps would essentially disappear in a highly nonuniform junction.
\section{CONCLUDING REMARKS}
A key premise of this paper is that the concept of coherent Josephson-like tunneling should be generalized to include other phenomena with quantum fluidic properties. Rather than treating the system as a massive object and computing the total Euclidean action or ``bounce," a generalized tunneling Hamiltonian matrix element is introduced to connect the original and emerging macrostates. The ``disconnectivity'' between macrostates, introduced by Leggett,~\cite{A.J.Leggett} is extremely small as in Josephson tunneling. This notion is further supported by examination of Schrieffer's real-space description of the BCS ground state:~\cite{J.R.Schrieffer} $\psi_0=\widehat{A}\varphi_0(\textbf{r}_1-\textbf{r}_2)\chi_{12}...\varphi_0(\textbf{r}_{N-1}-\textbf{r}_N)\chi_{N-1,N}$, where $\varphi_0$ is the relative coordinate wavefunction of a pair, $\chi$  is the corresponding spin function, and  $\widehat{A}$ is the antisymmetrization operator. As  $\psi_0$ evolves into a state, $\psi_{vp}$, containing a vortex-antivortex pair, each pair state individually evolves into a deformed state $\varphi_{vp}$ topologically connected to  $\varphi_0$. 

Beyond improving our understanding of superconducting flux vortex nucleation and dynamics, the concepts proposed here could have scientific impact in other areas, such as $\theta = \pi$ instabilities in spontaneous CP violation,~\cite{M.H.G.Tytgat} charge and spin density waves,~\cite{J.H.Miller.2011} topological insulators,~\cite{sRevD.78.054027,PRB83125119} and the quantum Hall effect.~\cite{PRB72035329} Finally, better understanding of the quantum behavior of flux solitons could potentially lead to topologically robust forms of quantum information processing.
\begin{acknowledgments}
  The authors acknowledge support by the State of Texas through the Texas Center for Superconductivity at the University of Houston. Additional support was provided by R21CA133153 from NIH (NCI) and by ARRA supplement: 3~R21~CA133153-03S1 (NIH, NCI).
\end{acknowledgments}



\begin{thebibliography}{References}
\bibitem{Josephson}	B.~D.~Josephson, Physics Letters \textbf{1}, 251 (1962).
\bibitem{3}	G.~Blatter, V.~B.~Geshkenbein, and V.~M.~Vinokur, Physical Review Letters \textbf{66}, 3297 (1991).
\bibitem{4}	P.~Ao and D.~J.~Thouless, Physical Review Letters \textbf{72}, 132 (1994).
\bibitem{5}	A.~Wallraff, A.~Lukashenko, J.~Lisenfeld, A.~Kemp, M.~V.~Fistul, Y.~Koval, and A.~V.~Ustinov, Nature \textbf{425}, 155 (2003).
\bibitem{E.M.Chudnovsky}	E.~M.~Chudnovsky, and J.~Tejada, \textit{Macroscopic Quantum Tunneling of the Magnetic Moment} (Cambridge University Press, Cambridge;  New York, 1998), p.~173.
\bibitem{Ihas1992} G.~G.~Ihas, O.~Avenel, R.~Aarts, R.~Salmelin, and E.~Varoquaux, Physical Review Letters  \textbf{69}, 327 (1992).
\bibitem{E.Grosfeld}	E.~Grosfeld and A.~Stern, Proceedings of the National Academy of Sciences \textbf{108}, 11810 (2011).
\bibitem{R.G.Mints}	R.~G.~Mints and I.~B.~Snapiro, Physical Review B \textbf{49}, 6188 (1994).
\bibitem{A.Widom}	A.~Widom and Y.~Srivastava, Physics Letters A \textbf{114}, 337 (1986).
\bibitem{T.D.Cohen}	T.~D.~Cohen and D.~A.~McGady, Physical Review D \textbf{78}, 036008 (2008).
\bibitem{Kalisky2009}   B.~Kalisky, J.~R.~Kirtley, E.~A.~Nowadnick, R.~B.~Dinner, E.~Zeldov, Ariando, S.~Wenderich,  H.~Hilgenkamp,  D.~M.~Feldmann, and K.~A.~Moler, Applied Physics Letters \textbf{94}, 202504 (2009).
\bibitem{D.V.Averin}	D.~V.~Averin and K.~K.~Likharev, Journal of Low Temperature Physics \textbf{62}, 345 (1986).
\bibitem{S.Coleman.1976}	S.~Coleman, Annals of Physics \textbf{101}, 239 (1976).
\bibitem{I.V.Krive}	I.~V.~Krive and A.~S.~Rozhavsky, Solid State Communications \textbf{55}, 691 (1985).
\bibitem{J.H.Miller}	J.~H.~Miller, C.~Ord\'{o}\~{n}ez, and E.~Prodan, Physical Review Letters \textbf{84}, 1555 (2000).
\bibitem{J.H.Miller.2011}	J.~H.~Miller, Jr., A.~I.~Wijesinghe, Z.~Tang, and A.~M.~Guloy, Physical Review Letters \textbf{108}, 036404 (2012).
\bibitem{H.Hilgenkamp}	H.~Hilgenkamp and J.~Mannhart, Reviews of Modern Physics \textbf{74}, 485 (2002).
\bibitem{J.H.Xu}	J.~H.~Xu, J.~L.~Shen, J.~H.~Miller,~Jr., and C.~S.~Ting, Physical Review Letters \textbf{73}, 2492 (1994).
\bibitem{D.Pekker}	D.~Pekker, G.~Refael, and P.~M.~Goldbart, Physical Review Letters \textbf{107}, 017002 (2011).
\bibitem{AppPhyLett69.3914} S.~K.~Tolpygo and M.~Gurvitch, Applied Physics Letters \textbf{69}, 3914 (1996).
\bibitem{A.A.Abrikosov}	A.~A.~Abrikosov,  (North-Holland, Elsevier Science Publishers, Amsterdam, 1988), p.~630.
\bibitem{S.R.Foltyn}	S.~R.~Foltyn, L.~Civale, J.~L.~MacManus-Driscoll, Q.~X.~Jia, B.~Maiorov, H.~Wang, and M.~Maley, Nature Materials \textbf{6}, 631 (2007).
\bibitem{J.Pearl}	J.~Pearl, Applied Physics Letters \textbf{5}, 65 (1964).
\bibitem{V.Selvamanickam}	V.~Selvamanickam~et~al., Applied Superconductivity, IEEE Transactions on \textbf{19}, 3225 (2009).
\bibitem{A.Maiti}	A.~Maiti and J.~H. Miller, Physical Review B \textbf{43}, 12205 (1991).
\bibitem{Z.Hermon}  Z.~Hermon, A.~Shnirman, and E.~Ben-Jacob, Physical Review Letters \textbf{74}, 4915 (1995).
\bibitem{R.P.Feynman}	R.~P.~Feynman, R.~B.~Leighton, and M.~Sands, \textit{The Feynman Lectures on Physics} (Addison-Wesley, Reading, Massachusetts, 1965), Vol.~III.
\bibitem{A.Grigorenko}	A.~Grigorenko, S.~Bending, T.~Tamegai, S.~Ooi, and M.~Henini, Nature \textbf{414}, 728 (2001).
\bibitem{prl77.3901} A.~E.~Koshelev, Physical Review Letters \textbf{77}, 3901 (1996).
\bibitem{Elion1993}   W.~J.~Elion, J.~J.~Wachters, L.~L.~Sohn, and J.~E.~Mooij, Physical Review Letters \textbf{71}, 2311 (1993).
\bibitem{J.Bardeen}	J.~Bardeen, Physical Review Letters \textbf{6}, 57 (1961).
\bibitem{C.B.Duke}	C.~B.~Duke, \textit{Tunneling in Solids} (Academic Press, New York, 1969), p.~xii~+~356.
\bibitem{J.Swihart}	J.~Swihart, J. Appl. Phys. \textbf{32}, 461 (1961).
\bibitem{R.D.Redwing}	R.~D.~Redwing, B.~M.~Hinaus, M.~S.~Rzchowski, N.~F.~Heinig, B.~A.~Davidson, and J.~E.~Nordman, Applied Physics Letters \textbf{75}, 3171 (1999).
\bibitem{Y.Y.Xue}	Y.~Y.~Xue, Z.~J.~Huang, P.~H.~Hor, and C.~W.~Chu, Physical Review B \textbf{43}, 13598 (1991).
\bibitem{X.Zhang}	X.~Zhang, S.~R.~Saha, N.~P.~Butch, K.~Kirshenbaum, J.~Paglione, R.~L.~Greene, Y.~Liu, L.~Yan, Y.~S.~Oh, K.~H.~Kim, and I.~Takeuchi, Applied Physics Letters \textbf{95}, 062510 (2009).
\bibitem{D.Dimos}	D.~Dimos, P.~Chaudhari, and J.~Mannhart, Physical Review B \textbf{41}, 4038 (1990).
\bibitem{F.Lombardi} F.~Lombardi, Z.~G.~Ivanov, P.~Komissinski, G.~M.~Fischer, P.~Larsson, and T.~Claeson, Applied Superconductivity \textbf{6}, 437 (1998).
\bibitem{C.Dark}	C.~Dark, S.~Speller, H.~Wu, A.~Sundaresan, Y.~Tanaka, G.~Burnell, and C.~R.~M.~Grovenor, Applied Superconductivity, IEEE Transactions on \textbf{15}, 2931 (2005).
\bibitem{Terpstra1995}	D.~Terpstra, R.~P.~J.~Ijsselsteijn, and H.~Rogalla, Applied Physics Letters \textbf{66}, 2286 (1995).
\bibitem{A.J.Leggett}	A.~J.~Leggett, Progress of Theoretical Physics Supplement \textbf{69}, 80 (1980).
\bibitem{J.R.Schrieffer}	J.~R.~Schrieffer, \textit{Theory of Superconductivity} (Benjamin Cummings, Reading, Massachusetts, 1964), p.~332.
\bibitem{M.H.G.Tytgat} M.~H.~G.~Tytgat, Physical Review D \textbf{61}, 114009 (2000).
\bibitem{sRevD.78.054027}	D.~Boer, and J.~K.~Boomsma, Physical Review D \textbf{78}, 054027 (2008).	
\bibitem{PRB83125119}	K.--T.~Chen and P.~A.~Lee, Physical Review B \textbf{83}, 125119 (2011).
\bibitem{PRB72035329}	A.~M.~M.~Pruisken, R.~Shankar, and N.~Surendran, Physical Review B \textbf{72}, 035329 (2005).

\end{thebibliography}
\end{document}